\documentstyle[12pt]{article}
\catcode`\@=11
\ifcase\@ptsize
 \font\tenmsy=msbm10
 \font\sevenmsy=msbm7
 \font\fivemsy=msbm5
 \font\teneu=eufm10
 \font\seveneu=eufm7
 \font\fiveeu=eufm5
\or
 \font\tenmsy=msbm10 scaled \magstephalf
 \font\sevenmsy=msbm8
 \font\fivemsy=msbm6
 \font\teneu=eufm10 scaled \magstephalf
 \font\seveneu=eufm8
 \font\fiveeu=eufm6
\or
 \font\tenmsy=msbm10 scaled \magstep1
 \font\sevenmsy=msbm8
 \font\fivemsy=msbm6
\font\teneu=eufm10   scaled \magstep1
\font\seveneu=eufm8
\font\fiveeu=eufm6
\fi
\newfam\msyfam
\textfont\msyfam=\tenmsy  \scriptfont\msyfam=\sevenmsy
  \scriptscriptfont\msyfam=\fivemsy
\def\Bbb{\ifmmode\let\next\Bbb@\else
 \def\next{\errmessage{Use \string\Bbb\space only in math mode}}\fi\next}
\def\Bbb@#1{{\Bbb@@{#1}}}
\def\Bbb@@#1{\fam\msyfam#1}
\newfam\eufam
\textfont\eufam=\teneu  \scriptfont\eufam=\seveneu
  \scriptscriptfont\eufam=\fiveeu
\def\frak{\ifmmode\let\next\frak@\else
 \def\next{\errmessage{Use \string\frak\space only in math mode}}\fi\next}
\def\frak@#1{{\frak@@{#1}}}
\def\frak@@#1{\fam\eufam#1}
\catcode`\@=12
\newtheorem{theorem}{Theorem}[section]

\renewcommand{\theequation}{\thesection.\arabic{equation}}
\renewcommand{\title}[1]{\large\bf
     #1\bigskip\medskip\\}
\renewcommand{\author}[1]{\large #1\\ \smallskip}
\newcommand{\address}[1]{{\normalsize\it #1\\}\bigskip}
\newcommand{\hs}[1]{\hspace*{#1cm}}
\newcommand{\vs}[1]{\vspace*{#1cm}}
\def\smat#1{\mbox{\small $\pmatrix{#1}$}}
\def\and{\;\;{\rm and}\;\;}
\def\disp{\displaystyle}
\def\be{\begin{eqnarray}}
\def\ee{\end{eqnarray}}
\def\no{\nonumber}
\def\half{\mbox{$\textstyle {1 \over 2}$}}
\def\3half{\mbox{\small$3\over 2$}}
\def\phalf{\mbox{\small$1\over 2$$p$}}
\def\h{\hspace*{0.5cm}}
\def\({\biggl(}
\def\){\biggr)}
\def\[{\mbox{$\biggl\lfloor$}}
\def\]{\mbox{$\biggr\rfloor$}}
\def\i{\rm i}
\def\T{\mbox{\boldmath { $T$}}}
\def\N{\mbox{\small${\cal N}$}}
\def\S{\mbox{\small${\cal S}$}}
\def\t{\mbox{\boldmath {$t$}}}
\def\scr{\scriptsize}
\def\lam{\lambda}
\def\disp{\displaystyle}
\def\I{\mbox{\boldmath $I$}}
\def\a{\mbox{$\alpha$}}
\def\U{\mbox{$\cal  A$}}
\def\e{\mbox{$\epsilon$}}
\def\b{\mbox{$\cal T$}}
\def\la{\mbox{${\ell}\!a$}}
\def\Le{\mbox{${\ell}\!e$}}
\def\lA{\mbox{${\ell}\!A$}}

\def\CC{{\Bbb C}}
\def\mo{\mbox{\boldmath $U$}}

\def\ade{\mbox{$A$--$D$--$E$ }}

\def\sqbox#1#2#3#4{\setlength{\unitlength}{0.0105in}
  \begin{picture}(20,20)(-#1,-#2)
   \put(0,0){\line(1,0){20}}
   \put(0,0){\line(0,1){20}}
   \put(20,20){\line(-1,0){20}}
   \put(20,20){\line(0,-1){20}}
   \put(7,7){\scr $#3$}
   \put(21,21){\scr $#4$}
\end{picture}}
\def\sqboxd#1#2#3{\setlength{\unitlength}{0.0105in}
  \begin{picture}(40,20)(-#1,-#2)
   \put(0,0){\line(0,1){20}}
   \put(40,0){\line(0,1){20}}
   \put(0,0){\line(1,0){40}}
   \put(0,20){\line(1,0){40}}
   \put(10,7){$\cdots$}
   \put(17,-20){\scr $#3$}
\end{picture}}
\def\sqqbox#1#2#3#4#5{\setlength{\unitlength}{0.0105in}
  \begin{picture}(20,40)(-#1,-#2)
   \put(0,0){\line(0,1){40}}
   \put(20,0){\line(0,1){40}}
   \put(0,0){\line(1,0){20}}
   \put(0,20){\line(1,0){20}}
   \put(0,40){\line(1,0){20}}
   \put(7,27){\scr $#3$}
   \put(7,7){\scr $#4$}
   \put(21,41){\scr $#5$}
\end{picture}}
\def\sqbbox#1#2#3#4{\setlength{\unitlength}{0.014in}
  \begin{picture}(20,20)(-#1,-#2)
   \put(0,0){\line(1,0){20}}
   \put(0,0){\line(0,1){20}}
   \put(20,20){\line(-1,0){20}}
   \put(20,20){\line(0,-1){20}}
   \put(8,8){\tiny $#3$}
   \put(21,21){\scr $#4$}
\end{picture}}

\def\Atwon#1#2#3#4{\setlength{\unitlength}{0.008in}%
\begin{picture}(#1,45)(0,0)
\multiput(0,0)(#1,0){2}{\line(0,1){30}}
\multiput(0,0)(0,30){2}{\line(1,0){#1}}
\put(#1,6){$\hs{0.3}#2$}\put(#3,-19){\scr$#4$}
\end{picture}}

\begin{document}
%\begin{flushright}Started:Nov 22, 1994\\ Today:\today\end{flushright}
\vspace{0.5cm}
\begin{center}
\vs{2}

\title{FUSION HIERARCHY AND FINITE-SIZE \\ CORRECTIONS OF  $U_q[sl(2)]$
INVARIANT
 \\ VERTEX MODELS WITH OPEN BOUNDARIES}

\author{Yu-kui Zhou\footnotemark\footnotetext{Email: {\sl
          ykzhou@mundoe.maths.mu.oz.au} before 30/04/95.}\footnotemark
    \footnotetext{
   On leave of absence from {\sl Institute of Modern Physics, Northwest
     University, Xian 710069, China}} }

\address{Mathematics Department, University of Melbourne,\\
          Parkville, Victoria 3052, Australia
\\and \\ Department of Mathematics, Australian National University
\\ Canberra, ACT 0200, Australia\footnotemark
%\\and \\ Institute of Modern Physics, Northwest University,\\ Xian 710069, P.
%%R. China
}
\footnotetext{Postal address after 01/05/95}
\end{center}

\begin{abstract}
The fused six-vertex models with open boundary conditions are studied.
The Bethe ansatz solution given by Sklyanin has been generalized to
the transfer matrices of the fused models.  We have shown that  the
eigenvalues of transfer matrices satisfy a group of functional relations,
which are the $su$(2) fusion rule held by the transfer matrices of the fused
models. The fused transfer matrices form a commuting family and also
commute with the quantum group $U_q[sl(2)]$. In the case of the
parameter $q^h=-1$ ($h=4,5,\cdots$) the functional relations
in the limit of spectral parameter $u\to \i\infty$ are truncated.
This shows that the $su$(2) fusion rule with finite level appears
for the six vertex model with the open boundary conditions. We have
solved the functional relations to obtain  the finite-size
corrections of the fused transfer matrices for low-lying excitations.
{}From the corrections
the central charges and conformal weights of underlying conformal
field theory are extracted. To see different boundary conditions
we also have studied the six-vertex model with a twisted boundary
condition.
\end{abstract}

\bigskip
hep-th/9502053; to appear in NPB.
\bigskip

\section{Introduction}
\setcounter{equation}{0}
In statistical mechanics  the commuting transfer matrix of
two-dimensional lattice models is often used. This is the case
because it shows obviously that the corresponding systems are
integrable and can be solved exactly. The commutativity of the
transfer matrices for periodic systems is easily derived from
the Yang-Baxter equation \cite{Yang,Baxter}. Recently, the
so-called open boundary condition has been introduced in the
study of two-dimensional lattice models. To have the commuting family
of transfer matrices for such systems we have to use the
Yang-Baxter equation for the bulk and the reflection equation
\cite{Cherednik} for  the boundaries.

The exactly solvable models with non-periodic boundaries have been
early studied in \cite{McWu,Bariev,Gaudin,Cardy:86,ABBBQ:87}.
After  Sklyanin's work on the algebraic Bethe ansatz of six-vertex
model with the open boundary condition \cite{Sklyanin},
there has been increasing interest in
exploring two-dimensional lattice models or integrable quantum chains
with the open boundary conditions \cite{KuSk:91}-\cite{PeBe:95}.
Recently, the boundary cross unitary has been derived in
\cite{GhZa:94} and a bootstrap approach has been developed to
two dimensional integrable field theory with boundary in
\cite{GhZa:94,FrKo:94} (see also \cite{SaBa:89,FeSa:94,Sasaki} for
related works). In \cite{JKKKM} a vertex operator approach
has been used to solve the semi-infinite $XXZ$ spin
with a boundary magnetic field.

The spin-$1/2$ $XXZ$ chain is the Hamiltonian limit of the
transfer matrix of six-vertex model. Similarly, the higher spin
$XXZ$ chains are the Hamiltonian limits of the fused
transfer matrices of six-vertex model \cite{KRS:81,ZaFa:80,KuRe:83,SAA}.
This corresponding relation also works for the model with the open
boundary conditions \cite{Sklyanin,MNR:90}. As the fusion of the
Boltzmann weights of six-vertex model \cite{KRS:81} so
the fusion procedure of the boundary matrices  has
been expressed in  \cite{MeNe:92}. The fused transfer matrices
with the open boundaries are well defined and form the commuting
families of the model. They satisfy a group of functional relations,
which can be shown by fusion procedure. For the Andrews, Baxter and
Forrester  critical solid-on-solid models \cite{ABF:84} it has shown
that the functional relations are useful to find the Bethe ansatz
solutions and the finite-size corrections of the fused
transfer matrices \cite{BaRe:89,KlPe:92}. But, for two dimensional
vertex models it seems not the case. The functional relations of the
fused transfer matrices of six-vertex model with periodic boundary are
given in \cite{KiRe:87} and have been shown that they are not
closed \cite{Paul:92}. However, the situation is very different for
the case of the six-vertex model with the open boundary conditions.
In this paper we study the six-vertex model with the open boundary
conditions. The fused transfer matrices of the model commute with the
quantum group $U_q[sl(2)]$. The eigenstates of the transfer
matrices can then be classified according to the spin $S^z=j$.
We focus on the interesting case of
$q^h=-1$ $(h=4,5,\cdots)$. We show that the functional relations of the
fused transfer matrices are truncated in braid limit. This important
fact allows the functional relations to be solved in thermodynamic
limit. We have found the finite-size corrections of the fused transfer
matrices for the low-lying excitations, which are the spin $S$ sectors
above the ground state. Therefore the central charges and the conformal weights
of
underlying conformal field theory have been extracted from the finite-size
corrections. The central charges are given by
\be
&&c\;=\;{3p\over p+2}-{6p\over h(h-p)}\; , \h h=4,5,\cdots
\ee
and the conformal weights are given by
\be
&&\Delta_{s,\nu}\;=\;{\( h-(h-p)s \)^2-p^2\over 4hp(h-p)} +
    {\nu(p-\nu)\over 2p(p+2)}
\h  s=1,3,\cdots\le h-1,
\ee
where  $p=1,2,\cdots$ is the fusion level and $\nu$ is the
integer determined by
\be
\nu=s-1-\[{s-1\over p}\]p \;.
\ee
Here the brackets $\left\lfloor x\right\rfloor$ denotes the greatest
integer less than or
equal to $x$.
For the case of $p=1$ these conformal spectra coincide with the
results of the spin-1/2 $XXZ$ chain with the open boundary
condition \cite{SaBa:89}. It is interesting to notice that for
generic $p$ these are the conformal spectra of the spin-$p/2$
$XXZ$ chain with the open boundary condition. In a similar way
we also have   obtained the finite-size
corrections of the fused transfer matrices of the six-vertex
model with a twisted boundary condition.

The outline of the paper is as follows. In subsection~1.1 and
subsection~1.2 we explain the six-vertex model and the open
boundary condition and recall the Bethe ansatz solution of the
$U_q[sl(2)]$ invariant transfer matrix of the six-vertex model.
In section~2 the functional equations of the fused transfer
matrices of the model are presented and the corresponding
Bethe ansatz of the transfer matrices are constructed.
In section~3 we carry out the procedure of calculating
the finite-size corrections of the fused transfer matrices
and extract the conformal spectra of underlying conformal
field theories. In section~4 a brief discussion is given.
Particularly, we discuss the non-$U_q[sl(2)]$ invariant
six-vertex model with boundaries.
In Appendices we describe the fused
Boltzmann weights of the six-vertex model and show
the functional equations directly by fusion.

\subsection{Six-vertex model}
The six-vertex model is one of solvable lattice models in two dimensional
statistical mechanics to prove tractable
\cite{Lieb,Sutherland,TakFad,Baxter}. It is ice-type model and has
six nonzero Boltzmann weights. These Boltzmann weights form a
four by four $R$-matrix
\be
R(z)=\smat{a(z)&0&0&0\cr
           0&b(z)&c(z)&0\cr
           0&c(z)&b(z)&0\cr
           0&0&0&a(z)} \; ,
\ee
where
\be
a(z)&=&zq-z^{-1}q^{-1} \no \\
b(z)&=&z-z^{-1}  \\
c(z)&=&q-q^{-1} \no
\ee
depending on a parameter $q$ and a spectral parameter $z\in \CC$.
This $R$ matrix acting on $\CC^2\otimes \CC^2$ solves the Yang-Baxter relation
\be
R^{12}(x/y)R^{13}(x/z)R^{23}(y/z)&=&R^{23}(y/z)R^{13}(x/z)R^{12}(x/y)\;.
\label{YBE}
\ee

For the twisted boundary
condition the row-to-row transfer matrix acting on a "quantum" space
$\CC^{2N}=\CC^2\otimes\cdots\CC^2$ is defined by
\be
\T_m(z)={\rm tr}\(\;q^{-m\sigma^z}\;\mbox{\boldmath $U$}(z)\)\; ,
\ee
where $m$ is an integer. It becomes the transfer matrix with the periodic
boundary condition for $m=0$ and for $m>0$ it is twisted.
The trace is taken in the "classical" space $\CC^2$ and
the monodromy matrix
\be
\mbox{\boldmath $U$}(z)&=&R^{c,N}(zq^{-\half})\cdots R^{c,2}(zq^{-\half})\
 R^{c,1}(zq^{-\half}) \label{U}
\ee
is a two by two matrix in the classical space $\CC^2$ denoted by $c$ and
with their elements acting on the quantum space $\CC^{2N}$.
%Where $\mu$ is a shift parameter.
The Yang-Baxter equation (\ref{YBE}) follows that the monodromy
matrix satisfies the quadratic relation
\be
R^{12}(x/y)\stackrel{1}\mo(x/z)\stackrel{2}\mo(y/z)&=&\stackrel{2}\mo(y/z)
\stackrel{1}\mo(x/z)R^{12}(x/y)\;
\ee
and the transfer
matrix forms a commuting family:
\be
\left[\T_m(z)\; ,\; \T_m(y)\;\right]=0\;.
\ee
This means that $\T_m(z)$ is a generating function of commuting operators
in quantum mechanics of one dimensional chains. Specially, the
Hamiltonian
\be
H_m&=&\sum_{n=1}^{N-1} \(\sigma_n^x\sigma_{n+1}^x+\sigma_n^y\sigma_{n+1}^y+
{q+q^{-1}\over 2}\sigma_n^z\sigma_{n+1}^z\)  \no \\
&&\h +\half\(q^{2m}\sigma^+_N\sigma_1^-+q^{-2m}\sigma^-_N\sigma_1^+
 +({q+q^{-1}})\sigma^z_N\sigma^z_1\)
\ee
of the quantum $XXZ$ spin-$\half$ chain is contained within this family.
These $\sigma_n^{x},\sigma_n^{y}$ and $\sigma_n^{z}$ are Pauli matrices
and $\sigma_n^{\pm}=\sigma_n^{x}\pm {\rm i}\sigma_n^{y}$.

The six-vertex model with the open boundary is given
by recalling the reflection equation \cite{Cherednik},
\be
R^{12}(x/y)K^1_{\pm}(x)R^{12}(xy)K^2_{\pm}(y)&=&
K^2_{\pm}(y)R^{12}(xy)K^1_{\pm}(x)R^{12}(x/y)\;.
\label{reflection}\ee
Sklyanin has shown in \cite{Sklyanin} that the transfer matrix
\be
\T(z)={\rm tr}\;K_+(zq^{1/2})\mbox{\boldmath $U$}(z)K_-(zq^{-1/2})
          {\mbox{\boldmath $U$}}^{-1}(z^{-1}) \label{T-o}
\ee
forms a commuting family
\be
\left[\T(z)\; ,\; \T(y)\;\right]=0\;.
\ee
Therefore it presents an integrable system with the boundary described
by the reflection matrices $K_{\pm}(z)$. Particularly, the matrices
$K_{\pm}(z)$ take the form \cite{KuSk:91}
\be
K_{\pm}(z)=\smat{z^{\mp}&0\cr 0&z^{\pm}} \; ,\label{openK}
\ee
then the integrable system possesses the quantum algebra
$U_q[sl(2)]$ symmetry, which means
\be
\T(y)\ S^{\pm}-S^{\pm}\ \T(y)\;=\;0\h {\rm and}\h
\T(y)\ S^{z}-S^{z}\ \T(y)\;=\;0      \label{T-q}
\ee
for any $y$. Here the operators $S^{\pm}$ and $S^{z}$
 of the quantum algebra $U_q[sl(2)]$
\be
S^{z}&=& \half(\sigma_1^z+\sigma_2^z+\cdots+\sigma_{N}^z) \\
S^{\pm}&=&\sum_{n=1}^N q^{(\sigma_1^z+\cdots+\sigma_{n-1}^z)/2}
     (\sigma_n^{\pm}/2)q^{-(\sigma_{n+1}^z+\cdots+\sigma_N^z)/2}\;
\ee
satisfy
\be
S^{z}S^{\pm}-S^{\pm}S^{z}\;=\;\pm S^{\pm} \h{\rm and}\h
S^{+}S^{-}-S^{-}S^{+}\;=\;[2S^{z}]_q
\ee
where
$$[x]_q={q^x-q^{-x}\over q-q^{-1}} \; . $$

The Hamiltonian corresponding to the integrable system
is the open $XXZ$ quantum spin-$\half$ chain \cite{ABBBQ:87}
\be
H&=&\sum_{n=1}^{N-1} \(\sigma_n^x\sigma_{n+1}^x+\sigma_n^y\sigma_{n+1}^y+
{q+q^{-1}\over 2}\sigma_n^z\sigma_{n+1}^z\)+{q-q^{-1}\over
2}\(\sigma^z_N+\sigma^z_1\)
\ee
The $U_q[sl(2)]$ invariance (\ref{T-q}) implies the quantum algebra
$U_q[sl(2)]$ is the "symmetric group" of the $XXZ$ chain \cite{PaSa:90}
\be
H\ S^{\pm}-S^{\pm}\ H\;=\;0\h {\rm and}\h
H\ S^{z}-S^{z}\ H\;=\;0      \label{H-q}
\ee
For generic values of $q$ the representations of $U_q[sl(2)]$ are
known to be equivalent to the ordinary su($2)$ representations
\cite{Rosso:88,Luztig:89}. The representation theory becomes more
complicated for the special case of $q^h=\pm 1$ (see
\cite{Luztig:89,ReSm:89,PaSa:90,ReTu:91,HSYY:91,JuKa:95}
for details).

\subsection{Bethe ansatz solution}
The Bethe ansatz equations and the eigenvalues of the six-vertex model
with the twisted or open boundary condition have been given using the
algebraic Bethe ansatz \cite{Baxter,Sklyanin,ABBBQ:87,TakFad}.
Set $z=e^{{\i}u}$
and $q=e^{{\i}\lam}$. We recall that the eigenvalues
$T_m(z)$ or $T(z)$ of the transfer matrices $\T_m(z)$ or $\T(z)$.
For the twisted boundary case the eigenvalues are given by
\be
T_m(u)&=&e^{-{\i}m\lam}\sin^N(u+\half\lam)
          \prod_{k=1}^M {\sin(u-v_k-\lam)\over \sin(u-v_m)} \no \\
     &&+\;e^{{\i}m\lam}\sin^N(u-\half\lam)
          \prod_{k=1}^M {\sin(u-v_k+\lam)\over \sin(u-v_m)}
\label{Tm(u)}\ee
and these $v_1,v_2,\cdots,v_M$ are the solutions of the Bethe ansatz
equations
\be
{\sin^N(v_k+\half\lam)\over \sin^N(v_k-\half\lam)}=
e^{2{\i}m\lam} \prod_{l\ne k}^M {\sin(v_k-v_l+\lam)\over \sin(v_k-v_l-\lam)}
\;.
\label{BAE1}\ee
For the open boundary case the eigenvalues are given by
\be
T(u)&=&{\sin(2u+\lam)\over \sin(2u)}\sin^{2N}(u+\half \lam)
          \prod_{m=1}^M {\sin(u-v_m-\lam)\sin(u+v_m-\lam)
            \over \sin(u-v_m)\sin(u+v_m)} \no \\
     &&+\;{\sin(2u-\lam)\over \sin(2u)}\sin^{2N}(u-\half \lam)
          \prod_{m=1}^M {\sin(u-v_m+\lam)\sin(u+v_m+\lam)
            \over \sin(u-v_m)\sin(u+v_m)}
\label{T(u)}\ee
and these $v_1,v_2,\cdots,v_M$ satisfy the Bethe ansatz equations
\be
\({\sin(v_m+\half \lam)\over \sin(v_m-\half \lam)}\)^{2N}=
\prod_{k\ne m}^M {\sin(v_m-v_k+\lam)\sin(v_m+v_k+\lam)
            \over \sin(v_m-v_k-\lam)\sin(v_m+v_k-\lam)} \; .
\label{BAE2}\ee

\section{Fused models and their functional equations}
\setcounter{equation}{0}

The fusion models can be built up by fusion \cite{KRS:81}
from the six vertex model. Suppose that $R_{(p,q)}(u)$ represents
the $R$-matrix of fused vertex (see Appendix. A) and then the
relevant monodromy matrix is defined by
\be
\mbox{\boldmath$U$}_{(p,q)}(u)&=&
R^{c,1}_{(q,p)}(u)R^{c,2}_{(q,p)}(u)\cdots,R^{c,N}_{(q,p)}(u)\; ,
\ee
where $p$ and $q$ are respectively the fusion levels for vertical
direction and horizontal direction of the square lattice and
$p,q=1,2,\cdots$. With the periodic boundary
condition the fused transfer matrices
\begin{equation}
\T^{(p,q)}(u)\;=\;{\rm tr}\;\mbox{\boldmath$U$}_{(p,q)}(u)
\end{equation}
commute
\be
\left[\T^{(p,q)}(u)\; ,\; \T^{(p,q')}(v)\;\right]=0\;
\ee
for each fusion level $p$ fixed and any $z$ and $y$. These
$q$ and $q'$ can stay in different levels. The transfer
matrices satisfy the following functional relations \cite{KiRe:87}
\be
&&\T^{(p,q)}(u)\T^{(p,1)}(u+q\lam)=\T^{(p,q+1)}(u)+f^p_{q-1}\T^{(p,q-1)}(u)\;,
\ee
where $\T^{(p,0)}(u)\;=\;\I$, the identity matrix,
and the $u$-dependent function $f^p_q$  is generated from the
antisymmetric fusion of the Boltzmann weights.
These relations are the su($2$) fusion rule. They
mean the relationship among the eigenvalues of the fused
transfer matrices. In other words, all eigenvalues $T^{(p,q)}$
of the fused transfer matrices are determined by the relations
with the initial solution $T^{(p,1)}$. In the following
subsection the similar idea is used to the transfer matrices
with open or twisted boundary condition. Let $T^{(p,q)}(u)$ be the
eigenvalues of the fused transfer matrices with the open boundary
condition or twisted boundary condition.
We prove the following theorems.
\begin{theorem}[$su(2)$ Fusion Hierarchy]\label{Ther-1}
Let us define
$$T^{(p,0)}\;=\;1 \h\h T^{(q)}_k\;=\;T^{(p,q)}(u+k\lam)\h
     \h f^p_{q}\;=\;f^p(u+q\lam)$$
\begin{equation}
     f^p(u) \;= \;\omega_1(u+\lam)\phi(u+\phalf\lam+\lam)
      \omega_2(u)\phi(u-\phalf\lam)\,
\end{equation}
where $\omega_1(u)$, $\omega_2(u)$ and $\phi(u)$ are given by
(\ref{omega-p})-(\ref{omega-o}) and (\ref{phi}).
Then the su($2$) fusion hierarchy follows
\be
\h T^{(q)}_0T^{(1)}_q= T^{(q+1)}_0  + f^p_{q-1}T^{(q-1)}_0
\label{Func-T}  \ee
for $q=1,2,\cdots$.
\end{theorem}
\begin{theorem}[$su(2)$ TBA]\label{Ther-2} If we define
\be
t^0_0&=&0 \;     \\
t^q_0&=&\disp{T^{(q+1)}_0T^{(q-1)}_1/ \disp\prod_{k=0}^{q-1}f^p_k }\; ,
\label{def-t}\ee
then it follows that $su(2)$ TBA equations
\be
t^q_0t^q_1=(1+t^{q+1}_0)(1+t^{q-1}_1)\;.\label{Func-t}
\ee
\end{theorem}

\subsection{Bethe ansatz for fused models}
The Bethe ansatz solutions (\ref{Tm(u)}) and (\ref{T(u)}) can be
written in the form of
\be
T(u)Q(u)&=&\omega_1(u) \phi(u+\half\lam)Q(u-\lam)+
    \omega_2(u) \phi(u-\half\lam)Q(u+\lam)
\ee
using Baxter's auxiliary matrix {\boldmath$Q$} which commutes with
the transfer matrix $\T(u)$. The eigenvalue $Q$ of the auxiliary matrix
{\boldmath$Q$} is given by
\be
Q(u)&=&\left\{ \begin{array}{ll}
\disp\prod_{m=1}^{M} \{ \sin(u-v_m) & \mbox{for twisted boundary}\vs{0.2}\\
\disp\prod_{m=1}^{M} \{ \sin(u-v_m)\sin(u+v_m)\}& \mbox{for open boundary}
\end{array}\right. \; \label{Q}
\ee
and $\phi(u)$ is given by
\be
\phi(u)&=&\sin^{\cal N}(u)
\label{phi}
\ee
with
\be
\N=\left\{\begin{array}{ll}
N &  \mbox{\rm for twisted boundary} \\
2N & \mbox{\rm for open boundary.}
\end{array}\right. \label{N}
\ee
The functions $Q(u)$ and $\phi(u)$ are not directly related to the
boundary. The boundary terms come
in the expression of eigenvalues through the factors $\omega_1(u)$
and $\omega_2(u)$
\be
\omega_1(u)\;=&\omega_2(u)^{-1}\;=\;e^{-{\i}\lam}&
   \mbox{for twisted boundary}\vs{0.2} \label{omega-p}\\
\omega_1(u)\;=&\omega_2(-u)\;=\;\disp{\sin(2u+\lam)\over \sin(2u)}&
  \mbox{for open boundary.}  \label{omega-o}
\ee
Here the simple case $m=1$ has been taken for the twisted boundary.

The fusion procedure \cite{KRS:81,MeNe:92}
(see \cite{Reshetikhin,DJKMO:87,ZPG:95,ZhPe:94,Zhou:94} for related works)
shows that the fused
transfer matrices $\T^{(p,1)}(u)$ can be constructed directly
from the unfused ones by applying the fusion projectors to the
"quantum" space (see Appendix. B). The procedure implies obviously
that the eigenvalue $T^{(p,1)}(u)$
for the transfer matrix $\T^{(p,1)}(u)$ has the following form
\be
\hs{-0.5}T^{(p,1)}(u)Q(u)=\omega_1(u) \phi(u+\phalf\lam)Q(u-\lam)+
    \omega_2(u) \phi(u-\phalf\lam)Q(u+\lam) \; .
\label{KuSk}\ee
The boundary conditions take the position in the "classical" space and
thus $\omega_1(u)$ and $\omega_2(u)$ are not effected by the fusion
procedure in "quantum" space. The function $Q(u)$ is however dependent on
the fusion level $p$. For example, the ground state corresponds
to take $M=pN/2$. The Bethe ansatz equations determining
all these $v_1,v_2,\cdots$ are given by setting
\be
T^{(p,1)}(v_k)=0\; .\label{BA}
\ee

To show  the su($2$) fusion hierarchy let
us use semi-standard Young tableaux \cite{KiRe:87,BaRe:89,KuSu:94,ZPG:95}.
Define
\be
{\sqbox{0}{-7}{\mbox{\scr $1$}}{}}^{k} & =
&\omega_2(u+k\lam)\phi(u+k\lam-\phalf\lam)
          {Q(u+k\lam+\lam)\over Q(u+k\lam)}\no \\
{\sqbox{0}{-7}{\mbox{\scr $2$}}{}}^{k} & =
&\omega_1(u+k\lam)\phi(u+k\lam+\phalf\lam)
         {Q(u+k\lam-\lam)\over Q(u+k\lam)} \label{box}
\ee
for a single Young tableau so that
\be
T^{(1)}_0 & = & \sqbox{0}{-7}{1}{}^{0} \; +\; \sqbox{0}{-7}{2}{}^{0} \;\; =\;\;
\sum \;\;\sqbox{0}{-7}{}{}^{0}
\ee
For a general one-row Young tableau, the numbers must not decrease moving to
the right along the row, e.g.
\be
{\sqbbox{0}{3}{\mbox{\small$1$}}{}
\sqbbox{0}{3}{\mbox{\small$1$}}{}
\sqbbox{0}{3}{\mbox{\small$2$}}{}
\sqbbox{0}{3}{\mbox{\small$2$}}{}
\sqbbox{0}{3}{\mbox{\small$2$}}{}}^{\;0}
\ee
Such a Young tableau denotes the product of the five labeled boxes defined
 by (\ref{box}) where it is understood that the relative shifts in the
spectral parameters are given \vs{0.2} by
\be
\sqbbox{0}{3}{\!\!\!\!\!\!u\!+\!4\lam}{}
\sqbbox{0}{3}{\!\!\!\!\!\!u\!+\!3\lam}{}
\sqbbox{0}{3}{\!\!\!\!\!\!u\!+\!2\lam}{}
\sqbbox{0}{3}{\!\!\!\!u\!+\!\lam}{}
\sqbbox{0}{3}{u}{\mbox{\scr$$}}^{\;0} \label{five-shifts}
\ee
and the zero superscript gives the shift in the most right box. Filling
the numbers $1$ and $2$ in this five-box Young tableau
according to the rule that the numbers must
not decrease moving to the right along the row, we get six numbered
Young tableaux. Then taking sum of these six Young tableaux with the
correct spectral parameter shifts (\ref{five-shifts}),
we obtain the eigenvalues
$T^{(5)}(u)$.

By a similar way the eigenvalues of the fused row transfer
matrix at level $q$ can be written  as
\be
T^{(q)}_{0} & = &T^{(p,q)}_{0}(u)\;=\; \sum  \begin{picture}(20,20)(-5,5)
    \put(0,0){${\underbrace
   {\sqbox{0}{3}{}{}
    \sqboxd{0}{3}{q}
    \sqbox{0}{3}{}{}}}^{\;0}$}
\end{picture} \label{Tq}
\ee

\noindent
where the number of terms in the sum is given by the dimension
of the irreducible representations of $su(2)$
\be
\mbox{dim}(q) = (q+1)\;.
\ee
For example, the fusion level $q=p=2$ case gives the
eigenvalues of transfer matrix of the $19$-vertex model
\be
T^{(2)}_{0} & = &\begin{picture}(180,20)(0,-5)
    \put(0,-13){\small${\sqbox{0}{3}{\mbox{\small$1$}}{}
\sqbox{0}{3}{\mbox{\small$1$}}{}}^{\;0}$}
    \put(75,-13){\small${\sqbox{0}{3}{\mbox{\small$1$}}{}
\sqbox{0}{3}{\mbox{\small$2$}}{}}^{\;0}$}
    \put(150,-13){\small${\sqbox{0}{3}{\mbox{\small$2$}}{}
\sqbox{0}{3}{\mbox{\small$2$}}{}}^{\;0}$}
\put(47,-5){$+$}\put(130,-5){$+$}
\end{picture}
\ee

\noindent
where it is understood that the relative shifts in the
arguments are given \vs{0.2} by
\be
\sqbbox{0}{3}{\!\!\!\!u\!+\!\lam}{}
\sqbbox{0}{3}{u}{\mbox{$$}}^{\;0}
\ee
It is straightforward to show that \vs{-0.1} set
\be
f^p(u) &:= & \sqqbox{0}{-17}{1}{2}{\mbox{\scr $0$}}
  \;:= \;
\omega_1(u+\lam)\phi(u+\phalf\lam+\lam)\omega_2(u)\phi(u-\phalf\lam)\;, \no
 \ee

\noindent
then $T^{(q)}_{0}$ given by (\ref{Tq}) satisfy (\ref{Func-T}).
This leads the theorem~\ref{Ther-1}. The proof also shows that
the fusion hierarchy is compatible with the su($2$) fusion rule
\be
&&\underbrace{\Atwon {180}{\otimes}{85}{q}} \hs{0.85}{\Atwon {30}{}{5}{}} \no
\\
&&\;=\;\underbrace{\Atwon {150}{\oplus}{55}{q-1}}\hs{0.85}
 \underbrace{\Atwon {210}{}{85}{q+1}}
\ee

To show theorem~\ref{Ther-2} let us consider the triple
$$ T^{(p,q)}_0(T^{(p,q-1)}_1T^{(p,1)}_q)=
(T^{(p,q)}_0T^{(p,1)}_q)T^{(p,q-1)}_1\; .$$
Inserting the fusion hierarchy into the terms in
parentheses this equation gives new functional
equations
\be
T^{(q)}_0T^{(q)}_1&=&\prod_{k=0}^{q-1}f^p_k  + T^{(q+1)}_0T^{(q-1)}_1\;,
\label{T-p}
\ee
which corresponds to the following su($2$) fusion rule
\be
&&\underbrace{\Atwon {180}{\otimes}{85}{q}}
  \hs{0.85}\underbrace{\Atwon {180}{}{85}{q}} \no \\
&&\;=\;\underbrace{\Atwon {150}{\otimes}{55}{q-1}}\hs{0.85}
 \underbrace{\Atwon {210}{\oplus\hs{0.2}{\mbox{\large$\phi$}}}{85}{q+1}}
\hs{0.85}
\ee

\noindent
Then it is easy to see the theorem~\ref{Ther-2} by rewriting
the fusion rule according to the definition of $t^q(u)$.

The functional equations (\ref{Func-T}) and (\ref{Func-t}) in form are
the same as the su($2$) functional equations of
\ade models \cite{BaRe:89,ZhPe:94}
and the dilute \ade models  \cite{Zhou:94}.

\subsection{Zeros and poles of eigenvalues }\label{sec:zeros}
The functional relations have been shown to be
very useful to calculate the
finite size corrections of the fused transfer matrices
\cite{KlPe:92,ZhPe:95}. To solve the fusion hierarchy (\ref{Func-T})
and (\ref{Func-t}) we need to know the distribution of zeros and
poles of these transfer matrices $\T^{(q)}$ and $\t^{(q)}$.
For $q=p$ these $\T^{(q)}$ possess the physical strip of the model.
Inside the strip the ground state eigenvalues $T^{(q)}$ do not
possess any zero apart from those which are imposed by the {\it fusion}
of the Boltzmann weights and the boundary. The zeros contributed
by the Boltzmann weights are of order $\N$ and those by
the boundary are only of order $1$. They list them as follows.
\be
{\rm zero}[T^{(p,q)}(u)]&=\;\mbox{$\emptyset$}\hs{3.7} &\mbox{for $q\le p$} \\
{\rm zero}[T^{(p,q)}(u)]&=\;\bigcup_{k=0}^{q-p-1}\{-k\lam-\phalf \lam\}^{N}
    &\mbox{for $q>p$.}
\ee
for the twisted boundary condition and
\be
{\rm zero}[T^{(p,q)}(u)]&=\;\bigcup_{k=0}^{q-2}\{-k\lam-\half\lam\}
     \hs{4.1}&\mbox{for $q\le p$} \\
{\rm zero}[T^{(p,q)}(u)]&=\;\bigcup_{k=0}^{q-2}\{-k\lam-\half\lam\}
    \bigcup_{k=0}^{q-p-1}\{-k\lam-\phalf\lam\}^{2N} &\mbox{for $q>p$.}
\ee
for the open boundary condition. The zeros and poles
of $t^{(q)}$ are determined by (\ref{def-t}). So we have
\be
{\rm (I)}\;\; q\le p-1\;: &&    \no\\
{\rm zero}[t^{(p,q)}(u)]&=& \mbox{$\emptyset$} \no\\
{\rm pole}[t^{(p,q)}(u)]&=&
     \bigcup_{k=0}^{q-1}\{-k\lam-\phalf\lam-\lam\}^N
    \bigcup_{k=0}^{q-1}\{\phalf\lam-k\lam\}^N   \\
{\rm (II)}\;\;  q= p\;:\h &&    \no\\
{\rm zero}[t^{(p,p)}(u)]&=& \{-\phalf\lam\}^N  \no\\
{\rm pole}[t^{(p,p)}(u)]&=& \bigcup_{k=0}^{p-1}\{-k\lam-\phalf \lam-\lam\}^{N}
\bigcup_{k=0}^{p-1}\{\phalf \lam-k\lam\}^{N}  \\
{\rm (III)}\;\;  q\ge p+1\;: &&    \no\\
\hs{-0.7}{\rm zero}[t^{(p,q)}(u)]&=& \mbox{$\emptyset$}\no\\
\hs{-0.7}{\rm pole}[t^{(p,q)}(u)]&=&
   \bigcup_{k=0}^{p-1}\{\phalf\lam-k\lam\}^N
  \bigcup_{k=1}^{p}\{-k\lam+\phalf\lam-q\lam\}^N
\ee
for the twisted boundary condition and
\be
{\rm (I)}\;\; q\le p-1\;: &&    \no\\
{\rm zero}[t^{(p,q)}(u)]&=& set_q \no\\
{\rm pole}[t^{(p,q)}(u)]&=&\{-q\lam-\half\lam\}\{\half\lam\}\no \\
  &&   \bigcup_{k=0}^{q-1}\{-k\lam-\phalf\lam-\lam\}^{2N}
    \bigcup_{k=0}^{q-1}\{\phalf\lam-k\lam\}^{2N}   \\
{\rm (II)}\;\;  q= p\;:\h &&    \no\\
{\rm zero}[t^{(p,p)}(u)]&=& \{-\phalf\lam\}^{2N}  \no\\
{\rm pole}[t^{(p,p)}(u)]&=&\{-p\lam-\half\lam\}\{\half\lam\}\no \\
  && \bigcup_{k=0}^{p-1}\{-k\lam-\phalf \lam-\lam\}^{2N}
       \bigcup_{k=0}^{p-1}\{\phalf \lam-k\lam\}^{2N}  \\
{\rm (III)}\;\;  q\ge p+1\;: &&    \no\\
{\rm zero}[t^{(p,q)}(u)]&=& \{\phalf\lam-q\lam\}^{2N}\no\\
{\rm pole}[t^{(p,q)}(u)]&=&
   \{-q\lam-\half\lam\}\{\half\lam\}\no \\
 && \bigcup_{k=0}^{p-1}\{\phalf\lam-k\lam\}^{2N}
  \bigcup_{k=1}^{p}\{-k\lam+\phalf\lam-q\lam\}^{2N}
\ee
for the open boundary condition, where $set_q=\{-\half\lam\}$ for $q=1$ and
$set_q=\emptyset$ for $q>1$.

The zeros or poles with order $1$ have less contribution than those of
order $\N$ when the system size $N$ becomes large. Especially, in the
thermodynamic limit $N\to\infty$ only these zeros or poles with order
$N$ are important.

\section{Functional relations in $N\to\infty$}
\setcounter{equation}{0}
The finite-size corrections for the eigenvalues $T^{(p)}$ can be
obtained by solving  functional relations (\ref{Func-T}) and
(\ref{Func-t}) in the physical strip,
\be
-\lam<{\rm Re}\;u+\phalf\lam<\lam  \; . \label{strip}
\ee
Denote the finite-size corrections of $T^{(p)}$
by $T^{(p)}_{\mbox{\tiny finite}}(u)$ and write
\be
T^{(p)}(u)=T^{(p)}_{\mbox{\tiny finite}}(u)
  T^{(p)}_{\mbox{\tiny bulk}}(u)\;.\label{def-finite}
\ee
The bulk and the surface energies determined by the
unitary conditions of $R$ and $K$ matrices and
satisfy
\be
T^{(p)}_{\mbox{\tiny bulk}}(u)T^{(p)}_{\mbox{\tiny bulk}}(u+\lam)
=\prod_{k=0}^{p-1}f^p_k
\ee
Inserting (\ref{def-finite}) into (\ref{T-p}) we find that
\be
T^{(p)}_{\mbox{\tiny finite}}(u)T^{(p)}_{\mbox{\tiny finite}}(u+\lam)
=1+t^{(p)}(u) \;.
\label{finite}\ee
So the finite-size  corrections for $T^{(p)}(u)$ are represented
by  the hierarchy $t^{(p)}(u)$ ($t$-system or $y$-system are also
called). In the following subsections the analytical treatment
of (\ref{finite}) and (\ref{Func-t}) is given. We will see that
the finite-size corrections in scaling limit are only dependent on
the braid asymptotics and bulk behavior of the functional relations.

\subsection{Nonlinear integral equations of real variable}
The Bethe ansatz equations (\ref{BA}) render $T^{(p)}(u)$ to be
analytic. Since all functions are $\pi$-periodic, the analyticity
domains for $T^{(p)}(u)$ are not unique. It is useful to introduce
functions of a real variable  by restricting the
eigenvalue functions to certain lines in the complex plane,
\be
&&\hs{1.2}{\b}^q(x):=T^{(q)}_{\mbox{\tiny finite}}
   \({{\i}\over \pi}x\lam +{p-q+1\over 2}\lam-\phalf\lam\)\; , \\
&&{\a}^q(x):=t^{(q)}\({{\i}\over \pi}x\lam +{p-q\over 2}\lam-\phalf\lam\)
\and {\U}^q(x):=1+{\it \a}^q(x)\; .
\ee
The functional relation (\ref{finite}) can then be rewritten in terms
of  the new functions as
\be
\b^p(x-\half\pi{\i})\b^p(x+\half\pi{\i})={\U}^{p}(x)\; .
\ee
For the ground state the functions $\U^{(p)}(x)$ and $\b^{(p)}(x)$
are {\sl
analytic, non-zero}\footnote{for those of order $\cal N$}
$\;$ in $-3\pi/2<{\rm Im}\;x<\pi/2$ and  possess {\sl constant}
asymptotics for ${\rm Re}\;x\to \pm \infty$ (ANZC), which can be seen
from the eigenvalues directly. Taking the logarithmic
derivative of the above equation and introducing Fourier transforms
\be
&&{B}^p(k)={1\over 2\pi}\int_{-\infty}^{\infty} dx\;[\ln\b^p(x)]'
    \;e^{-{\i} kx} \;,\no\\
&&\h [\ln\b^p(x)]'=\int_{-\infty}^{\infty} dk\;{B}^p(k) \;e^{{\i} kx} \;
\ee
with analogous equations for $\U^q$ and its Fourier transform $A^q$,
then we have
\be
{B}^p(k)={A^{p}\!(k)\over e^{(\pi/2)k}+e^{-(\pi/2)k}}\;.
\ee
Transforming back and defining the kernel $k(x)$
\be
k(x):= {1\over 2\pi\cosh(2x)} \;,
\ee
we are able to express $\b^q$ in terms of $\U^q$,
\be
\ln\b^p=k*\ln \U^p +C^p\;, \label{b}
\ee
where $C^p$ are integration constants. The convolution $f*g$ of
two functions $f$ and $g$ is defined by
\be
(f*g):=\int^\infty_{-\infty}f(x-y)g(y)\;dy=
\int^\infty_{-\infty}g(x-y)f(y)\;dy\; .
\label{convolution}\ee
In case of the low-lying excitations states we have to take care of zeros
in the analyticity strips so that the simple ANZC properties hold. The result
(\ref{b}) is still correct if we change integration path ${\cal L}$ so that
$\b^p(x)$ has an ANZC area and Cauchy theorem can be applied like the
discussion in  \cite{KlPe:92}. The integration constants in (\ref{b})
can be evaluated from the asymptotics
of $\U^q$ and $\b^q$. In this limit (\ref{b}) becomes
\be
\ln\b^q_\infty=\half\ln \U^q_\infty +C^q\;.
\ee
It can be seen that the constants are just the multiple of $\i\pi$
and do not contribute to the $1\over N$ corrections.

The $\U^q$ can be solved from the hierarchy from (\ref{Func-t}),
which can be rewritten in terms of $\a^q$ as
\be
\a^q(x-\half\pi{\i})\a^q(x+\half\pi{\i})={\U}^{q-1}(x){\U}^{q+1}(x)\; .
\ee

According to section~\ref{sec:zeros} the analyticity strip (\ref{strip})
for $t^{(p)}(u)$ contains a zero of order $\N$ at $u=-\phalf\lam$ and
a pole  of order $\N$ at $u=-\phalf\lam+\lam$ or $u=-\phalf\lam-\lam$.
All other functions $t^{(b,q)}$ are  analytic and non-zero in
their analyticity strips $-\phalf\lam-\lam< u<-\phalf\lam+\lam$.
We introduce finite-size correction terms $l^q(x)$ by writing $\a^q(x)$ as
\be
\a^q(x)=\left\{\begin{array}{ll}
 l^q(x)\; , & q\not=p \\
\tanh^{\mbox{\scr$\cal N$}}(\half x)l^q(x)\; , & q=p \;.
\end{array}\right. \label{a-l}
\ee
The factor $\tanh^{\mbox{\scr$\cal N$}}(\half x)$ gives
the right zero and poles and  all the functions $l^q(x)$
therefore are ANZC in $-\pi<{\rm Im}\;x<\pi$.
They satisfy the functional equations
\be
l^q(x-\half\pi{\i})l^q(x+\half\pi{\i})=\U^{q-1}(x)\U^{q+1}(x)\; .
\ee
Again applying Fourier transforms to the logarithmic derivative of the
equations
and then integrating the equations back we obtain the nonlinear integral
equations
\be
\ln\a^q=\ln\e^q+k*\ln\U^{q-1}+k*\ln\U^{q+1}+D^q\; , \label{a}
\ee
where
\be
\e^q(x):=\left\{ \begin{array}{ll}        \label{e}
1 \; , & q\not=p \\
\tanh^{\mbox{\scr$\cal N$}}(\half x)\; , & q=p \;.
\end{array}\right.
\ee
$D^q$ are the integral constants. For the same reason we have to take
care of the extra zeros in the analyticity strips so that the ANZC
is held in (\ref{a}).

\subsection{Finite-size correction and scaling limit}
The information of finite-size corrections can be extracted from the
nonlinear integral equations (\ref{a}) and (\ref{b}).
The system size $N$ enters the nonlinear equations (\ref{a}) through
(\ref{e}). The function $\e^p$ has three asymptotic regimes with transitions
in scaling regimes when $x$ is of the order of $-\ln N$ or $\ln N$.
We suppose that $\a^q$ and $\U^q$ scale similarly. So in the following
scaling limits,
\be
&&e^q_{\pm}(x):=\lim_{{N}\to \infty}\e^q\(\pm(x+\ln {\N})\) \; ,   \no \\
&&a^q_{\pm}(x):=\lim_{{N}\to \infty}\a^q\(\pm(x+\ln {\N})\) \; ,\\
&&A^q_{\pm}(x):=\lim_{{N}\to \infty}\U^q\(\pm(x+\ln {\N})\)=1+a^q_\pm(x)\; .\no
\label{scaling}\ee
In this scaling limits, (\ref{a}) takes the form
\be
\la^q=\Le^q+k*\lA^{q-1}+k*\lA^{q+1}+D^q \; ,\label{a-L}
\ee
where we use the abbreviations
\be
&& \la^q(x):=\ln a^q(x) \; ,\h \lA^q(x):=\ln A^q(x) \; , \no\\
&& \Le^q(x):=\left\{\begin{array}{ll}
    0\; , & q\not=p \;, \\
  -2 e^{-x}\; , &q\not=p \;  \end{array}\right. \label{Le}
\ee
and suppress the subscripts $\pm$.
The transfer matrix $\b^p(x)$ in $N\to\infty$ now becomes
\be
&&\ln\b^p(x)=(k*\ln\U^p)(x) \no \\
&&={1\over 2\pi}\int^\infty_{-\ln {\cal N}}\({\ln\U^p(y+\ln {\N})\over
 \cosh(x-y-\ln {\N})}+{\ln\U^p(-y-\ln {\N})\over
 \cosh(x+y+\ln {\N})}\)\;dy  +{ o\!}\({1\over {\N}}\) \no\\
&&={e^x\over {\N}\pi}\int^\infty_{-\infty} e^{-y}\lA^p_+(y)\;dy
+{e^{-x}\over {\N}\pi}\int^\infty_{-\infty} e^{-y}\lA^p_-(y)\;dy
+{ o\!}\({1\over {\N}}\)  \no \\
&&={2\cosh x\over {\N}\pi}\int^\infty_{-\infty} e^{-y}
\lA^p(y)\;dy+{ o\!}\({1\over {\N}}\)  \;.\label{b-N}
\ee
Above equation converges and actually can be
evaluated explicitly with the help of the dilogarithmic function
\be
L(x)=-\int_0^x dy\;{\ln (1-y)\over y} +\half \ln x\ln (1-x) .
\ee
Multiplying the derivative of (\ref{a-L}) with $\lA^q$, and (\ref{a-L})
itself with $(\lA^q)'$, taking the difference, summing over $q$, and finally
integrating we find
\be
&&\sum_{q}^{}\int_{-\infty}^{\infty}[(\la^q)'\lA^q-\la^q(\lA^q)']\;
dx\no\\&&\h=\sum_{q}^{}\int_{-\infty}^{\infty}
[(\Le^q)'\lA^q-(\Le^q-D^q)(\lA^q)']\; dx \; ,  \label{la-lA}
\ee
where the sum is over all fusion levels $q$ and
the contribution of the kernel cancel due to the symmetry
\be
k(-x)=k(x)\; . \label{kernel-sym}
\ee
Then inserting (\ref{Le}) into the right-hand side and
integrating the left-hand side of (\ref{la-lA}),
we are able to obtain
\be
2\int_{-\infty}^\infty e^{-y}\lA^p(y)\;dy=
-\sum_{q}^{}L\({1\over A^q}\)\rule[-15pt]{0.2mm}{35pt}_{-\infty}^\infty+
\half\sum_{q}^{}D^q\lA^q\;\rule[-15pt]{0.2mm}{35pt}_{-\infty}^\infty
 \label{integral}\ee
where the constants $D^q$ are given in terms of
\be
D^q=\la^q-\half \lA^{q-1}-\half \lA^{q+1} \; \label{D}
\ee
in asymptotics $x\to\infty$.
\subsection{Asymptotics and bulk behavior}\label{Asmpsec}
The nonlinear integral equations (\ref{a-L}) can be easily
solved for the limit $x\to\pm\infty$ and
\be
\lam={\pi\over h}\h h=3,4,\cdots\;.
\ee
For different $h$ it corresponds to different models. The equation
(\ref{integral}) shows that these asymptotic solutions are enough
to obtain the finite-size corrections of the transfer matrix
$T^{(p)}(u)$. Before discuss the asymptotic solutions it is
useful to observe that
\be
T^{p,h-1}(\pm\i\infty)=0\;  \h\mbox{for $\lam={\pi/ h}\;$,}
\ee
which can be easily seen from the eigenvalues (\ref{KuSk}) and the
theorem~2.1.

It is obvious to see that the asymptotics $x\to\infty$ corresponds to the
braid limit of $u\to\pm{\i}\infty$. In this limit (\ref{Func-t})
reduces to
\be
(t^{(q)}_\infty)^2=(1+t^{(q-1)}_\infty)(1+t^{(q+1)}_\infty) \; .\label{rec}
\ee
This equation in turn means
\be
2\la^q=\lA^{q-1}+ \lA^{q+1}+D^q
\ee
in terms of the functions $a^q$. Where the constants
$D^{q}$ can be zero or non-zero because the
different branches can be taken for the dilogarithmic functions
in the nonlinear integral equations.
%%%%%%%%%%%%%%%%%%%%%%%%%%%%%%%%%%%%%%%%%%%%%%%%%%%%%%%%%%%%%%

To solve the $t^{(q)}_\infty$ let us write $t^{(1)}_\infty$ as
$$t^{(1)}_\infty={\sin(3\theta)\over\sin(\theta)} $$
with the parameter $\theta$ to be determined. The recursion relation
(\ref{rec}) implies
\be
&&t^{(q)}_\infty={\sin(q\theta)\sin\( (q+2){\theta}^{}\)
\over \sin^2\theta}\; \no\\
&&\hs{.1}t^{(q)}_\infty+1= {\sin^2\( (q+1)\theta\)\over \sin^2\theta}\;
\ee
for all $q=1,2\cdots$. This solution have to be consistent with the
braid limit of $T^{(q)}(u)$. To fix the constant parameter
$\theta$ let us consider the
"ground" state $M=\phalf N$,
\be
\lim_{{\rm Im}u\to\pm\infty}T^{(1)}(u)/\phi(u)=2\cos\({\pi\over h}\)\;.
\ee
By the relation
\be
&&{\sin(3\theta)\over\sin(\theta)}\;=\;
t^{(1)}_\infty\;=\;\lim_{{\rm Im}u\to\pm\infty}{T^{(2)}_0\over f^p_0}\;=\;\no
\\
&&\lim_{{\rm Im}u\to\pm\infty}{T^{(1)}_0T^{(1)}_1\over f^p_0}-1\;=\;
4\cos^2\({\pi\over h}\)-1
\ee
we have
\be
\theta=\lam={\pi\over h}\;.
\ee
Moreover the special value of $\theta$ leads the closure condition
\be
t^{(h-2)}_\infty=0 \;. \label{clo}%\h\h \mbox{for  $q=0\;\;$ mod. $h-2$}
\ee
For the sector $j=\phalf N-M$ we have to modify $\theta$ to be
\be
\theta=m\lam={m\pi\over h}\;
\ee
where $m=2j+1=1,3,\cdots\le h-1$. For the periodic case $m=1,2,3,\cdots,h-1$.

In the limit of $x\to-\infty$ $t^{(q)}$ can be considered as the bulk
behavior in large $N$. According to section~\ref{sec:zeros} the
analyticity strip for $t^{(p)}(u)$ contains a zero of order $N$ at
$u=-\phalf\lam$ and poles of order $N$ at $u=-\phalf\lam\pm\lam$.
All other functions $t^{(q)}$ are analytic and non-zero in their
analyticity strips in $-\phalf\lam-\lam< u<\phalf\lam+\lam$.
For large $N$ the leading bulk behavior to the $t^{(q)}$ we find
that
\be
t^{(q)}_{\rm bulk}(u)=\left\{ \begin{array}{ll}
 \mbox{constant ,}                            & q\not=p \; ,\\
 \mbox{constant}\(\tan({1\over 2}h u)\)^{\cal N}\; , & q=p \; .
\end{array} \right. \label{bulk-t}\ee
The constants are fixed by the functional equations (\ref{Func-t})
and can be calculated similarly to the asymptotics of these $t^{(q)}$.
Like $A_L$ model \cite{KlPe:92}, it is easy to see that the limit
\be
\lim_{x\to-\infty}\lim_{N\to\infty}t^{(p)}\sim \lim_{x\to-\infty}
\exp{(-2e^{-x})}\to 0\;.
\ee
Therefore the functional equations (\ref{Func-t}) are divided into
two parts and we find the constants for $1\le q\le p-1$
\be
&&t^{(q)}_{\rm bulk}={\sin(q\sigma)\sin\( (q+2){\sigma}^{}\)
\over \sin^2\sigma}\; \no\\
&&\hs{.1}t^{(q)}_{\rm bulk}+1= {\sin^2\( (q+1) \sigma\)\over \sin^2\sigma}\;,
\ee
where
\be
\sigma={m^,\pi\over p+2}    \hs{1} m^,=1,2,\cdots,p+1 \; .
\ee
Similarly, for $p+1\le q\le h-3$ we suppose
\be
&&\hs{-1}t^{(q)}_{\rm bulk}={\sin\((q-p)\tau\)\sin\( (q-p+2){\tau}\)\over
\sin^2\tau}
\no\\&&t^{(q)}_{\rm bulk}+1= {\sin\( (q-p+1) \tau\)\over \sin^2\tau} \;
\ee
with
\be
&& \tau={m^{,,}\pi\over h-p}    \;,
\ee
which is consistent with the closure condition (\ref{clo}).
The eigen-spectra of the transfer matrices  have only one "quantum"
number $M$, which is related
to the braid limit. These $m^,$ and $m^{,,}$ can not be free parameters.
In \cite{KlPe:92} an interpolate method is applied to compute
the finite size corrections of transfer matrices for ABF models.
This follows that the exponents $m'$
from the bulk behavior are no longer independent,
\be
m'=m-m^{,,} +2 n +1\; \label{mmm}
\ee
with the integer $n$ given by
\be
 n=\[{m-m^{,,}\over p}\] \;,
\ee
where the brackets $\left\lfloor x\right\rfloor$ denotes the greatest
integer  less than or equal to $x$. Here for the periodic boundary
system we need two parameters $m,m^{,,}$ and
we suppose that $m^{,,}=1,2,\cdots,h-p-1$ and $m'$ is given by (\ref{mmm}).
For the largest eigenvalue (or the ground state), the appropriate choices are
$m=m^,=m^{,,}=1$ and $m,m^,,m^{,,}>1$ give the low-lying excited states.
The open boundary systems possess the $U_q[sl(2)]$ invariance. According to
the study of $XXZ$-chain \cite{SaBa:89} we modify (\ref{mmm}) to be
\be
m'=m+2 n \h \h  n=\[{m-1\over p}\] \;
\ee
or suppose that $m^{,,}=1$
and then $m^,$ is determined by $m$. The low-lying excited states
are given by $m,m^{,}>1$.

The solution $t^{(p)}_{\rm bulk}(u)$ is given by
\be
t^{(p)}_{\rm bulk}(u)t^{(p)}_{\rm bulk}(u+\lambda)&=&(1+t^{(p+1)}_{\rm bulk})
        (1+t^{(p-1)}_{\rm bulk})  \no \\
&=& 16 \cos^2\sigma \cos^2 \tau
\ee
Thus we find lastly that
\be
t^{(p)}_{\rm bulk}(u)=\pm 4 \cos\sigma\cos\tau\(\tan({1\over 2}h u)\)^{\cal N}
\;.
\ee

\subsection{Central charge and conformal weights}
The finite-size corrections are only dependent on the braid and bulk
limits of the models. In these limits the functional relations are truncated
and therefore the sum in (\ref{integral}) can be replaced with the finite
sum
\be
2\int_{-\infty}^\infty e^{-y}\lA^p(y)\;dy=
-\sum_{q=1}^{h-3}L\({1\over A^q}\)
 \rule[-15pt]{0.2mm}{35pt}^{\infty}_{-\infty\;\; .}+
 \half\sum_{q=1}^{h-3}D^{q}
 \lA^{q}\;\rule[-15pt]{0.2mm}{35pt}_{-\infty}^\infty
 \label{integra2}\ee
%%%%%%%%%%%%%%%%%%%%%%%%%%%%%%%%%%%%%%%%%%%%%%%%%%%%%%%%%%%%%%%%%%%%%%%%%%%
where the constants $D^{(b,q)}$ can be zero or non-zero because the
different branches can be taken for the dilogarithmic functions in
the nonlinear integral equations. The choice of branches have to
give the right finite size corrections. Simply taking $D^{(b,q)}=0$
is consistent with the asymptotics solutions given in
subsection~\ref{Asmpsec}. To take nonzero
$D^{(b,q)}$ we have to single out the right branches of the dilogarithm
for the asymptotic solutions of the equations, which have been shown
for ABF models in \cite{KlPe:92}.
%%%%%%%%%%%%%%%%%%%%%%%%%%%%%%%%%%%%%%%%%%%%%%%%%%%%%%%%%%%%%%

The following useful dilogarithm identity has been established by
Kirillov~\cite{Kirillov:93}. Consider the functions
\be
 y^{(q)}\!(j,r):={\sin(q+2)\varphi\;\sin(q\varphi)\over
          \sin^2(\varphi)}\; ,
 \h 1\le b\le n-1,\h 1\le q\le r \label{su(n)2}
\ee
with
\be
\varphi={(1+j)\pi\over 2+r }\h 0\le j\le r\;.
\ee
It is obvious that they are the asymptotic solutions of the  functions
equations (\ref{a-L}) with $r=h-2$ or the bulk behavior of the functions
equations with $r=p$ and $r=h-2-p$.  Then the following dilogarithmic
function identity holds,
\be s(j,r)&:=&
\sum_{q=1}^{r}L({1\over 1+y^{(q)}\!(j,r)})
      \label{dilogarithm-s}      \no\\
&=&L(1)\({3r\over 2+r}-{6j(j+2)\over 2+r} +\;6j\) \;.   \label{s}
\ee

In terms of the dilogarithm function the finite-size corrections
(\ref{b-N}) are expressed as
%%%%%%%%%%%%%%%%%%%%%%%%%%%%%
\be
\ln\b^{(p)}(x)={\cosh x\over \N\pi}\(\;\;\sum_{q=1}^{h-3}
  L\({1\over A^{q}}\) \rule[-15pt]{0.2mm}{35pt}_{-\infty}^\infty+
 \half\sum_{q=1}^{h-3}D^{q}
 \lA^{q}\;\rule[-15pt]{0.2mm}{35pt}_{-\infty}^\infty\;\;\)
 +{ o\!}\({1\over {\N}}\)\label{bb}
 \;.
\ee
%%%%%%%%%%%%%%%%%%%%%%%%%%%%
Here we  take the  case of $D^q\not=0$. Note that the nonlinear
integral equations (\ref{a-L}) including the closure condition
(\ref{clo}) and their solutions presented in
subsection~\ref{Asmpsec} are the same as ones of the ABF models
studied in \cite{KlPe:92}. Therefore we can calculate the
finite size corrections in the same way.
Similarly to \cite{KlPe:92,Kirillov:93}, it can be shown
that in terms of the functions $s(j,r)$ the finite-size corrections
(\ref{bb}) for the open boundary systems can be expressed as
\be
\ln\b^{(p)}(x) &=&{\pi\cosh x\over 6\N} \(
                          s(0,h-2-p)+s(\nu,p)-s(m-1,h-2)\no\\
&&\hs{2}-{6(1-m)(p+1-m)+6\nu(p-\nu)\over p}\)
 +{ o\!}\({1\over {\N}}\) \;,
\label{excitation1}
\ee
where $\nu$ is an unique integer determined by
\be \nu=m-1-\[{m-1\over p}\]p \;.
\ee
Inserting (\ref{s}) into (\ref{excitation1}) we have the
finite-size correction
\be
\ln\b^p(x)={\pi\over 6\N}\(c-24\Delta_{m,\nu}\)\cosh x
 +{ o\!}\({1\over N}\), \label{b-cd}
\ee
where the center charges $c$ are given by
\be
c&=&{3p\over p+2}-{6p\over h(h-p)}\;  \label{c}
\ee
and the conformal weights are given by
\be
\Delta_{m,\nu}&=&{\( h-(h-p)m \)^2-p^2\over 4hp(h-p)} +
    {\nu(p-\nu)\over 2p(p+2)} \label{delta}
\hs{1.5}\ee
where $p=1,2,\cdots,h-2$, $m=1,3,\cdots\le h-1$.
Taking into account the geometrical factor $\cosh x=\sin(uh)$ and $\sinh x
={\i}\cos (hu)$ we obtain the expression of finite-size correction
to the energy given in \cite{BCN:86,Affleck:86}. For the special case
of $p=2$ and $m-1=even$ the same result has been claimed in
\cite{PaSa:90}.

For the twisted boundary case we have
\be
\ln\b^{(p)}(x) &=&{\pi\cosh x\over 6\N} \(
                          s(m^{,,}-1,h-2-p)+s(\nu,p)-s(m-1,h-2)\no\\
&&\hs{2}-{6(m^{,,}-m)(p+m^{,,}-m)+6\nu(p-\nu)\over p}\)
 +{ o\!}\({1\over {\N}}\) \no \\
&=&{\pi\over 6\N}\(c-24\Delta_{m^{,,},\nu,m}\)\cosh x
 +{ o\!}\({1\over N}\)\;,
\label{excitation2}
\ee
where $\nu$ is an unique integer determined by
\be \nu=m-m^{,,}-\[{m-m^{,,}\over p}\]p \;.
\ee
The conformal weights are given by the following
standard expression
\be
\Delta_{t,\nu,s}\;=\;{\( ht-(h-p)s \)^2-p^2\over 4hp(h-p)} +
    {\nu(p-\nu)\over 2p(p+2)} \label{Kac}
\ee
with
\be \nu=s-t-\[{s-t\over p}\]p \;.
\ee
given in \cite{DJKMO:87,KlPe:92}.
Comparing with the standard expression we find that
$t=1$,$s=1,3,\cdots\le h-1$ for
the fused six-vertex models with
the open boundary or $t=1,2,\cdots,h-p-1$,$s=1,2,\cdots,h-1$ for
the fused six-vertex models with the twisted boundary.

\section{Discussion }
We have constructed the functional relations among the fused transfer
matrices of the six-vertex model with the open boundary conditions.
The fusion procedure shows that the fusion level of the model can be
any positive integers and therefore
the fusion hierarchy of the six-vertex model
with periodic boundary conditions   is infinite.
The functional relations correspond to the su($2$)
fusion rule without truncation. It means that the underlying
algebra su($2$) has infinite level. This theory has the central
charge $\disp{3p/(2+p)}$.
By open boundaries, however, the
situation can be dramatically changed. The functional relations
for the six vertex model with the open boundary in
braid limit are truncated for $q^h=e^{h\i\lam}=-1$. So the su($2$)
fusion rule with a finite level appears again for
the model. This shows that the open boundary lower the central
charges to be less that $\disp{3p/(2+p)}$.

   The central charges of underlying conformal field theories for
the fused six-vertex models with the open boundary condition
have been found to be the same as those for the
fused ABF models. The open boundaries ensure $U_q[sl(2)]$ invariance
and also play the role of the charge at infinite in the Feigin
Fuchs construction \cite{FeFu:82}. The conformal weights of the
models now take only subset of the Kac formula (\ref{Kac}) or
$\Delta_{s,\nu}=\Delta_{1,\nu,s}$. The quantum number $s$ from
the braid solutions of the model is specified by the symmetric
algebra $U_q[sl(2)]$. That means that the spin  $S$ enters the
calculation through the asymptotics of inversion identity
hierarchies $t^q$. Suppose that $S^z=j$. Then we have $s=2j+1$
\cite{PaSa:90}.
Specially, we have known that
the partition function is a single form
in Virasoro characters \cite{SaBa:89,Cardy:89,PaSa:90}.

For the models with the twisted boundary we have found the
same central charges and the conformal weights. But $U_q[sl(2)]$
is no longer the "symmetric group". With the similar analysis
to \cite{PaSa:90} we can find that the states of $T_m(u)$
with different $m$ are mapped on each other by the
$U_q[sl(2)]$. We need two integers $s$ and $t$
for the representations of Virasoro algebras. As we know
the partition function is a sesquilinear form in Virasoro
characters.

The six-vertex model with the boundary specified by the reflection
matrices $K_+(u)=K(-u-\lam,\xi_+)$ and $K_-(u)=K(u,\xi_-)$
\be
K(u)=\smat{\sin(u+\xi)&0\cr 0&\sin(-u+\xi)}\;
\ee
is the original model studied by Sklyanin \cite{Sklyanin}.
This model does not possess $U_q[sl(2)]$ invariance.
In the case of $\xi_+ +\xi_-=\disp{\pi\over h}$
we can  solve the finite size corrections similarly and
thus the similar conformal spectra (\ref{c}) and (\ref{delta}).
One fact to see this is to quickly check the braid limit
of the transfer matrices, which are the same as those of
the model with the open boundary condition (\ref{openK}).

The conformal spectra given in this paper is for the
fusion hierarchy and therefore is more general.
For the case of $p=1$ the underlying  model is the unfused
six-vertex model. The conformal spectra coincides with
that of spin-$\half$  $XXZ$ chain with open
boundary \cite{ABBBQ:87,PaSa:90}. The conformal spectra
for the general fusion level $p$  gives the spectra
of spin-$\phalf$  $XXZ$ chains with the open boundaries.

The analytic method to calculate the finite size corrections
of transfer matrix  by solving the functional equations
has been described for study of the ABF models \cite{KlPe:92}.
Here we have
generalized the method to find the finite size corrections
of transfer matrix with open boundaries. We like to mention
that there is other method available to calculate
the finite size corrections
of transfer matrix with open boundaries, which generalizes
the method described in \cite{KBP:91,Zhou}.

\section*{Acknowledgements} This research has been supported by the
Australian Research Council. The author also thanks B. Y. Hou and P. A.
Pearce for discussions.

%\newpage
\bigskip
{\bigskip\rm\large\bf\noindent  Appendix~A: Fused weights of the six-vertex
model}
\setcounter{equation}{0}
\renewcommand{\theequation}{A.\arabic{equation}}

\noindent
We give the explicit expression  for the fused weights of
the six-vertex model in this section. Let $Y_p$ be the projector on the space
of symmetric tensor in $\CC^{2p}$,
\be
Y_p&=&{1\over p!}(P^{1,p}+\cdots+P^{p-1,p}+I)\cdots(P^{1,2}+I) \no \\
P^{i,j}&=& R^{i,j}(0) /\sin(\lam)\;.
\ee
The fused weight $R_{p,q}(u)\in \CC^{2s+1}\otimes\CC^{2s+1}$ is defined
by
\be
R_{(p,q)}(u)&=&Y_qR_{(p,q)}(u-q\lam+\lam)\cdots
R_{(p,2)}(u-\lam)R_{(p,1)}(u)Y_q \\
R_{(p,j)}(u)&=&Y_pR^{1,j}(u)R^{2,j}(u+\lam)\cdots R^{p,j}(u+p\lam-\lam)Y_p
\ee
The derivation of the fused $R$ matrix is straightforward and it is
$(p+1)^2$ by $(p+1)^2$ matrix with the following elements
\be
%% FOLLOWING LINE CANNOT BE BROKEN BEFORE 80 CHAR
\hs{-0.2}R_{(p,q)}(u)_{i,j}^{k,l}=C(u)\sum_{n}\(F(n)_{i,j}^{k,l}F(n,u)_{i,j}^{k,l}\)
 \h i,j,k,l=-s,-s+1.\cdots,s\ee
\be
C(u)\;=\;\prod_{m=0}^{2s-1}\(\sin^{-1}(2s\lam-m\lam)
     \prod_{n=1}^{2s-1}\sin(u+n\lam-m\lam)\)
\ee
\be
\hs{-0.6}F(n)_{i,j}^{k,l}=\prod_{m=1}^{l\!-\!j\!+\!n}
          {\sin(s\!+\!i\!+\!n\!-m\!+\!1)\lam\over \sin(m\lam)}
  \prod_{m=1}^{n} {\sin(s\!-\!i\!-\!m\!+\!1)\lam\over \sin(m\lam)}
  \prod_{m=1}^{s\!-\!j\!-\!1} {\sin(s\!-\!j\!-\!m)\lam\over \sin(2s\!-\!m)\lam}
\hs{-0.4}\ee
\be
\hs{-0.5}F(n,u)_{i,j}^{k,l}= \prod_{m=1}^{s-l-n}
   {\sin(u\!+\!s\lam\!-\!i\lam\!-\!n\lam\!-\!m\lam\!+\!\lam)\over\sin(m\lam)}
  \prod_{m=1}^{s+j-n} {\sin(u\!+\!i\lam\!+\!j\lam\!-\!m\lam\!+\!\lam)\over
\sin(m\lam)}
\ee
where $s=\phalf$ and the sum over $n$ is from $max(0,j-l)$
to $min(s+j,s-l,s-i)$. The matrix $R_{p,1}(u)$ can be written as
$2$ by $2$ matrix
\be
R_{(p,1)}(u)=\prod_{n=1}^{p-1}\sin(u\!+\!n\lam)\sum_{m=0}^{p-1}
\smat{\sin(u\!+\!p\lam\!-\!m\lam)e_{m,m}&\sin(p\lam\!-\!m\lam)e_{m\!+\!1,m}\cr
      \sin(m\lam)e_{m-1,m}&\sin(u+m\lam)e_{m,m}}
\ee
where the matrix $e_{i,j}\in End(\CC^{2s+1})$ has only non-zero entry
$(i,j)$ which is $1$.

{\bigskip\rm\large\bf\noindent  Appendix~B: Eigenvalue problem of $T^{(p,1)}$}
\setcounter{equation}{0}
\renewcommand{\theequation}{B.\arabic{equation}}

\noindent
Here we would like to show the Bethe ansatz solutions of $T^{(p,1)}$.
The transfer matrix $T^{(p,1)}$
\be
\T_{(p,1)}(z)={\rm tr}\;K_+(zq^{1/2})\mbox{\boldmath
$U$}_{(p,1)}(z)K_-(zq^{-1/2})
          {\mbox{\boldmath $U$}}^{-1}_{(p,1)}(z^{-1})
\label{B1}\ee
with the monodromy matrix
\be
\mbox{\boldmath $U$}_{(p,1)}(z)&=&R^{c,N}_{(1,p)}(z)\cdots
 R^{c,2}_{(1,p)}(z)\ R^{c,1}_{(1,p)}(z)
\ee
is given by the fusion in the "quantum" space. Removing the zeros generated
from fusion and replacing $u$ by $u-\phalf\lam$ the fused $R$ matrix reads
\be
R_{(1,p)}(u)=\sum_{m=0}^{p-1}
%% FOLLOWING LINE CANNOT BE BROKEN BEFORE 80 CHAR
\smat{\sin(u\!+\!\phalf\lam\!-\!m\lam)e_{m,m}&\sin(p\lam\!-\!m\lam)e_{m\!+\!1,m}\cr
    \sin(m\lam)e_{m-1,m}&\sin(u\!-\!\phalf\lam\!+\!m\lam)e_{m,m}}
\ee
which can be rewritten as
\be
R_{(1,p)}(z)=\smat{zq^{{\cal S}^z}-z^{-1}q^{-{\cal S}^z}&(q-q^{-1}){\cal
S}^-\cr
                   (q-q^{-1}){\cal S}^+&zq^{-{\cal S}^z}-z^{-1}q^{{\cal S}^z}}
\label{L}\ee
where the operators $\S^z,\S^\pm$ are generators of
$U_q[sl(2)]$
\be
q^{{\cal S}^z}\S^\pm q^{-{\cal S}^z}=q^\pm\S^\pm \h
\S^{+}\S^{-}-\S^{-}\S^{+}\;=\;[2\S^{z}]_q
\ee
The matrix (\ref{L}) is just the $L$-matrix used in \cite{KuSk:91}.
Therefore the Bethe ansatz solutions of $\T_{(p,1)}(z)$ defined in
(\ref{B1}) have been given exactly in  \cite{KuSk:91}
(also see \cite{MeNe:91}), which are
the equations (\ref{KuSk}) and (\ref{BA}).

{\bigskip\rm\large\bf\noindent  Appendix~C: Functional equations from fusion
procedure }
\setcounter{equation}{0}
\renewcommand{\theequation}{C.\arabic{equation}}

\noindent
In this section we explain how the functional equations of
the fused transfer matrices
come out from fusion procedure. As a simple example, we only show
that
\be
 \T^{(1)}_0\T^{(1)}_1= \T^{(2)}_0  + f^p_0 I \label{C1}
\ee
graphically. For this purpose let us represent the $R-$ and $K-$ matrices by
\be
\setlength{\unitlength}{0.0115in}%
\begin{picture}(120,60)(60,345)
\put(120,375){\vector( 1, 0){ 60}}
\put(150,345){\vector( 0, 1){ 60}}
\put(141,363){\scr$u$}
\put( 50,372){$R^{12}(u)$ $=$}
\end{picture}
\setlength{\unitlength}{0.0115in}%
\begin{picture}(201,60)(30,714)
\put(108,714){\line( 1, 1){ 30}}
\put(138,744){\vector(-1, 1){ 15}}
\put(258,774){\line(-1,-1){ 30}}
\put(228,744){\vector( 1,-1){ 15}}
\put(123,759){\line(-1, 1){ 15}}
\put(243,729){\line( 1,-1){ 15}}
\put( 36,735){$;$}
\put( 57,740){$K_-(u)\;=\hs{0.65}u$}
\put(150,735){$;$}
\put(167,740){$K_+(u)\;=\hs{0.4}u$}
\end{picture}
\ee
Therefore we can represent the Yang-Baxter
equation (\ref{YBE})  by
\be\setlength{\unitlength}{0.0105in}%
\begin{picture}(183,63)(69,717)
\put( 81,765){\vector( 1, 0){  3}}
\put(147,765){\vector(-1, 0){  6}}
\put(135,774){\vector( 0,-1){  6}}
\put( 75,765){\line( 1, 0){ 15}}
\put( 90,765){\line( 1,-1){ 30}}
\put(120,735){\line( 1, 0){ 30}}
\put( 75,735){\line( 1, 0){ 12}}
\put( 87,735){\line( 6, 5){ 35.5}}
\put(122.5,765){\line( 1, 0){ 30}}
\put(135,780){\line( 0,-1){ 60}}
\put(243,765){\vector(-1, 0){  3}}
\put(177,765){\vector( 1, 0){  6}}
\put(189,774){\vector( 0,-1){  6}}
\put(249,765){\line(-1, 0){ 15}}
\put(234,765){\line(-1,-1){ 30}}
\put(204,735){\line(-1, 0){ 30}}
\put(249,735){\line(-1, 0){ 12}}
\put(237,735){\line(-6, 5){ 35.5}}
\put(201.5,765){\line(-1, 0){ 30}}
\put(189,780){\line( 0,-1){ 60}}
\put(162,744){$=$}
\put( 96,753){\scr$u\!\!+\!\!v$}
\put(138,768){\scr$u$}
\put(128,737){\scr$v$}
\put(192,738){\scr$u$}
\put(180,768){\scr$v$}
\put(210,753){\scr$u\!\!+\!\!v$}
\put( 69,762){\small$3$}
\put( 69,732){\small$1$}
\put(190,717){\small$2$}
\put(129,717){\small$2$}
\put(252,762){\small$1$}
\put(252,732){\small$3$}
\end{picture}\label{YB}
\ee
the reflection equation (\ref{reflection}) by
\be
\setlength{\unitlength}{0.0085in}%
\begin{picture}(381,135)(111,540)
\put(120,675){\line( 2,-3){ 60}}
\put(180,585){\line(-2,-3){ 30}}
\put(117,648){\line( 2,-1){ 63.600}}
\put(180,615){\line(-2,-1){ 63.600}}
\put(180,675){\line( 0,-1){135}}
\put(213,633){\line( 2,-1){ 63.600}}
\put(276,600){\line(-2,-1){ 63.600}}
\put(216,540){\line( 2, 3){ 60}}
\put(276,630){\line(-2, 3){ 30}}
\put(276,675){\line( 0,-1){135}}
\put(198,606){$=$}
\put(225.7,555){\vector( 3, 4){  0}}
\put(234,579){\vector( 3, 2){0}}
\put(144,598){\vector( 3, 2){0}}
\put(165,561){\vector( 1, 1){0}}
\put(111,579){\small$1$}
\put(144,540){\small$2$}
\put(224,540){\small$2$}
\put(207,570){\small$1$}
\put(153,594){\scr$u\!\!+\!\!v$}
\put(252,612){\scr$u\!\!+\!\!v$}
\put(135,636){\scr$u\!\!-\!\!v$}
\put(231,570){\scr$u\!\!-\!\!v$}
\put(174,567){\scr$v$}
\put(174,621){\scr$u$}
\put(270,588){\scr$u$}
\put(270,639){\scr$v$}
\put(380.2,555){\vector(3, -4){0}}
\put(372,579){\vector(3, -2){0}}
\put(462,597.4){\vector(3, -2){0}}
\put(441,561){\vector(1, -1){0}}
\put(330,675){\line( 0,-1){135}}
\put(426,675){\line( 0,-1){135}}
\put(486,675){\line(-2,-3){ 60}}
\put(426,585){\line( 2,-3){ 30}}
\put(489,648){\line(-2,-1){ 63.600}}
\put(426,615){\line( 2,-1){ 63.600}}
\put(393,633){\line(-2,-1){ 63.600}}
\put(330,600){\line( 2,-1){ 63.600}}
\put(390,540){\line(-2, 3){ 60}}
\put(330,630){\line( 2, 3){ 30}}
\put(495,579){\small$1$}
\put(462,540){\small$2$}
\put(386,542){\small$2$}
\put(399,570){\small$1$}
\put(408,606){$=$}
\put(342,612){\scr$u\!\!+\!\!v$}
\put(435,597){\scr$u\!\!+\!\!v$}
\put(465,639){\scr$u\!\!-\!\!v$}
\put(351,588){\scr$u\!\!-\!\!v$}
\put(429,567){\scr$v$}
\put(429,621){\scr$u$}
\put(330,636){\scr$v$}
\put(333,588){\scr$u$}
\put(297,603){or}
\end{picture}\label{RYB}
\ee
and the unitary relation $R^{1,2}(u)R^{1,2}(-u)\sim I$ by
\be
\setlength{\unitlength}{0.0115in}%
\begin{picture}(144,36)(137,372)
\put(126,405){\line( 1, 0){ 24}}
\put(150,405){\line( 1,-1){ 30}}
\put(180,375){\line( 1, 0){ 30}}
\put(210,375){\line( 1, 1){ 30}}
\put(240,405){\line( 1, 0){ 21}}
\put(261,405){\line(-1, 0){  3}}
\put(126,375){\line( 1, 0){ 24}}
\put(150,375){\line( 1, 1){ 30}}
\put(180,405){\line( 1, 0){ 27}}
\put(207,405){\line( 6,-5){ 35.5}}
\put(242.7,375){\line( 1, 0){ 21}}
\put(192,375){\vector( 1, 0){ 0}}
\put(189,405){\vector( 1, 0){ 0}}
\put(156,387){\scr$u$}
\put(210,387){\scr$-\!u$}
\put(117,372){\small$2$}
\put(117,399){\small$1$} \put(300,387){$\sim$}
\put(330,387){$I$}
\end{picture}\label{I}\ee
The transfer matrix $T(u)$ with open boundary (\ref{T-o}) is then
represented by
\be
\setlength{\unitlength}{0.0125in}%
\begin{picture}(165,60)(60,735)
\put(150,750){\vector( 1, 0){  6}}
\put(105,735){\vector( 0, 1){ 60}}
\put(120,735){\vector( 0, 1){ 60}}
\put(195,735){\vector( 0, 1){ 60}}
\put(180,735){\vector( 0, 1){ 60}}
\multiput(138,765)(8.00000,0.00000){4}{\makebox(0.4444,0.6667){\tenrm .}}
\put( 75,789){\line( 0,-1){ 45}}
\put(225,786){\line( 0,-1){ 45}}
\put(129,780){\line(-1, 0){ 39}}
\put( 90,780){\line(-1,-1){ 15}}
\put( 75,765){\line( 1,-1){ 15}}
\put( 90,750){\line( 1, 0){120}}
\put(210,750){\line( 1, 1){ 15}}
\put(225,765){\line(-1, 1){ 15}}
\put(210,780){\line(-1, 0){ 84}}
\put(140,780){\vector(-1, 0){0}}
\put( 99,741){\scr$u$}
\put(174,741){\scr$u$}
\put(198,771){\scr$u$}
\put(108,771){\scr$u$}
%\put( 60,768){\scr$-\!u\!-\!\lam$}\put(219,771){\scr$u$}
\put(22,764){\scr$K_+(u\!+\!\half\lam)$}
\put(230,764){\scr$K_-(u\!-\!\half\lam)$}
\end{picture}
\ee

Let us consider $\T^{(1)}_0\T^{(1)}_1$
\be\setlength{\unitlength}{0.0125in}%
\begin{picture}(186,120)(48,675)
\put(105,735){\vector( 0, 1){ 60}}
\put(120,735){\vector( 0, 1){ 60}}
\put(195,735){\vector( 0, 1){ 60}}
\put(180,735){\vector( 0, 1){ 60}}
\multiput(138,765)(8.00000,0.00000){4}{\makebox(0.4444,0.6667){\tenrm .}}
\put(150,690){\vector( 1, 0){  6}}
\put(105,675){\vector( 0, 1){ 60}}
\put(120,675){\vector( 0, 1){ 60}}
\put(195,675){\vector( 0, 1){ 60}}
\put(180,675){\vector( 0, 1){ 60}}
\multiput(138,705)(8.00000,0.00000){4}{\makebox(0.4444,0.6667){\tenrm .}}
\put(129,720){\line(-1, 0){ 39}}
\put( 90,720){\line(-1,-1){ 15}}
\put( 75,705){\line( 1,-1){ 15}}
\put( 90,690){\line( 1, 0){120}}
\put(210,690){\line( 1, 1){ 15}}
\put(225,705){\line(-1, 1){ 15}}
\put(210,720){\vector(-1, 0){ 84}}
\put( 75,792){\line( 0,-1){114}}
\put(225,795){\line( 0,-1){114}}
\put(129,780){\line(-1, 0){ 39}}
\put( 90,780){\line(-1,-1){ 15}}
\put( 75,765){\line( 1,-1){ 15}}
\put( 90,750){\line( 1, 0){120}}
\put(210,750){\line( 1, 1){ 15}}
\put(225,765){\line(-1, 1){ 15}}
\put(210,780){\vector(-1, 0){ 84}}
\put(159,750){\vector( 1, 0){  6}}
\put(234,762){\scr$K_-(u\!+\!\half\lam)$}
\put(22,762){\scr$K_+(u\!+\!\3half\lam)$}
\put(195,771){\scr$u\!+\!\lam$}
\put(234,702){\scr$K_-(u\!-\!\half\lam)$}
\put(22,702){\scr$K_+(u\!+\!\half\lam)$}
\put( 99,681){\scr$u$}
\put(198,711){\scr$u$}
\put( 96,747){\small$b$}
\put( 96,717){\small$a$}
\put(168,741){\scr$u\!+\!\lam$}
\end{picture}\ee
Inserting the identical operator  into the position
$a,b$ and using the unitary condition (\ref{I}) and
the Yang-Baxter equation (\ref{YB}) (the spectra
parameter $u$ is shifted to be $u-\half\lam$ in
the monodromy matrix (\ref{U})), we are
able to obtain
\be\setlength{\unitlength}{0.0125in}%
\begin{picture}(192,105)(39,525)
\put( 75,630){\line( 0,-1){105}}
\put(225,630){\line( 0,-1){105}}
\put( 75,570){\line( 1, 1){ 30}}
\put(105,600){\line( 1, 0){ 90}}
\put(195,600){\line( 1,-1){ 30}}
\put(225,570){\line(-1,-1){ 15}}
\put(210,555){\line(-1, 0){120}}
\put( 90,555){\line(-1, 1){ 15}}
\put(111,540){\vector( 0, 1){ 90}}
\put(126,540){\vector( 0, 1){ 90}}
\put(186,540){\vector( 0, 1){ 90}}
\put(171,540){\vector( 0, 1){ 90}}
\multiput(138,585)(8.00000,0.00000){4}{\makebox(0.4444,0.6667){\tenrm .}}
\put(144,555){\vector( 1, 0){  6}}
\put(144,570){\vector( 1, 0){  6}}
\put(144,615){\vector(-1, 0){  3}}
\put(147,600){\vector(-1, 0){  3}}
\put( 75,600){\line( 1,-1){ 30}}
\put(105,570){\line( 1, 0){ 90}}
\put(195,570){\line( 1, 1){ 30}}
\put(225,600){\line(-1, 1){ 15}}
\put(210,615){\line(-1, 0){120}}
\put( 90,615){\line(-1,-1){ 15}}
\put(22,597){\scr$K_+(u\!+\!\3half\lam)$}
\put(22,552){\scr$K_+(u\!+\!\half\lam)$}
\put(231,552){\scr$K_-(u\!-\!\half\lam)$}
\put(231,597){\scr$K_-(u\!+\!\half\lam)$}
\put(105,546){\scr$u$}
\put(186,606){\scr$u\!+\!\lam$}
\put(120,594){\scr$u$}
\put(126,561){\scr$u\!+\!\lam$}
\put( 78,588){\scr$-\!2\!\lam\!-\!2\!u$}
\put(198,573){\scr$\phantom{+\lam}\!\!\!2\!u$}
\put(102,600){\small$a$}
\put( 96,615){\small$c$}
\end{picture}\label{C8}
\ee
The fully symmetric and antisymmetric operators are
given by the $R$ matrix,
\be
Y_2&=&\half(I+P^{1,2})\;=\;D^+ R^{1,2}(\lam)  \\
Y^-&=&\half(I-P^{1,2})\;=\;D^- R^{1,2}(-\lam)
\ee
where
\be
D^\pm=\smat{\sin^{-1}(2\lam)&0&0&0\cr
          0&\pm\half\sin^{-1}(\lam)&0&0\cr
          0&0&\pm\half\sin^{-1}(\lam)&0\cr
          0&0&0&\sin^{-1}(2\lam)}
\ee
Then inserting the identical operator  into the position
$a,c$ of (\ref{C8}) and using $Y_2+Y^-=I$ and the fusion
of the $R$ matrices
\be\setlength{\unitlength}{0.0125in}%
\begin{picture}(357,60)(160,285)
%\begin{picture}(357,60)(60,285)
%\put( 60,303){\vector( 1, 0){ 60}}%\thicklines
%\put( 75,285){\vector( 0, 1){ 33}}
%\put(105,285){\vector( 0, 1){ 33}}
%\put( 98,295){\scr$u$}
%\put(55,294){\scr$u\!-\!\lam$}
%\put(180,297){\thicklines\vector( 0, 1){ 39}}
%\put(165,315){\vector( 1, 0){ 33}}
%\put(173,307){\scr$u$}
%\put( 75,315){\vector( 1, 1){ 30}}
%\put(105,315){\vector(-1, 1){ 30}}
%\put(86.2,335){\scr$\lam$}
%\put(138,309){$\sim$}
\put(300,285){\vector( 0, 1){ 60}}
\put(285,300){\vector( 1, 0){ 30}}
\put(285,330){\vector( 1, 0){ 30}}
\put(294,291){\scr$u$}
\put(280,322){\scr$u\!+\!\lam$}
\put(399,294){\vector( 0, 1){ 39}}
\put(384,312){\thicklines\vector( 1, 0){ 33}}
\put(392,304){\scr$u$}
\put(315,330){\vector( 1,-1){ 30}}
\put(315,300){\vector( 1, 1){ 30}}
\put(366,309){$\sim$}
\put(333,312){\scr$\lam$}
%\put(228,309){and}
\end{picture}\ee
and the $K$ matrices
\be\setlength{\unitlength}{0.0125in}%
\begin{picture}(246,96)(27,696)
\put(243,780){\line( 0,-1){ 75}}
\put(273,768){\thicklines\line(-1,-1){ 30}}
\put(243,738){\thicklines\line( 1,-1){ 30}}
\put(267,714){\vector( 1,-1){0}}
\put(178,735){\scr$K^{(1,2)}_+(u\!+\!\half\lam)$}
\put( 66,792){\line( 0,-1){ 96}}
\put( 96,792){\line(-1,-1){ 30}}
\put( 66,762){\line( 1,-1){ 60}}
\put(126,732){\line(-1,-1){ 30}}
\put( 96,702){\line(-1, 1){ 30}}
\put( 66,732){\line( 1, 1){ 45}}
\put( 93,759){\vector(-1,-1){0}}
\put( 78,774){\vector(-1,-1){0}}
\put(126,732){\vector( 1, 1){0}}
\put(124,704){\vector( 1,-1){0}}
\put(120,711){\scr$\lam$}
\put( 87,741){\scr$-\!2\!\lam\!-\!2u$}
\put(15,726){\scr$K_+(u\!+\!\half\lam)$}
\put(15,756){\scr$K_+(u\!+\!\3half\lam)$}
\put(156,735){$\sim$}
\end{picture}\ee
\be\setlength{\unitlength}{0.0125in}%
\begin{picture}(192,90)(36,537)
\put(222,618){\line( 0,-1){ 75}}
\put(192,618){\thicklines\line( 1,-1){ 30}}
\put(222,588){\thicklines\line(-1,-1){ 30}}
\put(228,582){\scr$K^{(1,2)}_-(u\!-\!\half\lam)$}
\put(204,606){\vector(-1, 1){0}}
\put( 36,627){\line( 1,-1){ 30}}
\put( 66,597){\line( 1,-1){ 30}}
\put( 96,567){\line(-1,-1){ 30}}
\put( 36,597){\line( 1, 1){ 30}}
\put( 66,627){\line( 1,-1){ 30}}
\put( 96,597){\line(-1,-1){ 39}}
\put( 96,627){\line( 0,-1){ 90}}
\put( 81,552){\vector( 1, 1){0}}
\put( 66,567){\vector( 1, 1){0}}
\put( 36,597){\vector(-1,-1){0}}
\put( 36,627){\vector(-1, 1){0}}
\put( 36,609){\scr$\lam$}
\put(100,564){\scr$K_-(u\!-\!\half\lam)$}
\put(100,594){\scr$K_-(u\!+\!\half\lam)$}
\put(53,579){\scr$\phantom{+\!\lam}2\!u$}
\put(162,579){$\sim$}
\end{picture}
\ee
The term involved by $Y_2$ in (\ref{C8}) gives the transfer
matrix with the open boundary of fusion level $2$
\be
\setlength{\unitlength}{0.0125in}%
\begin{picture}(165,60)(60,735)
\put(150,750){\vector( 1, 0){  6}}
\put(105,735){\vector( 0, 1){ 60}}
\put(120,735){\vector( 0, 1){ 60}}
\put(195,735){\vector( 0, 1){ 60}}
\put(180,735){\vector( 0, 1){ 60}}
\multiput(138,765)(8.00000,0.00000){4}{\makebox(0.4444,0.6667){\tenrm .}}
\put( 75,789){\line( 0,-1){ 45}}
\put(225,786){\line( 0,-1){ 45}}
\put(129,780){\thicklines\line(-1, 0){ 39}}
\put( 90,780){\thicklines\line(-1,-1){ 15}}
\put( 75,765){\thicklines\line( 1,-1){ 15}}
\put( 90,750){\thicklines\line( 1, 0){120}}
\put(210,750){\thicklines\line( 1, 1){ 15}}
\put(225,765){\thicklines\line(-1, 1){ 15}}
\put(210,780){\thicklines\line(-1, 0){ 84}}
\put(140,780){\vector(-1, 0){0}}
\put( 99,741){\scr$u$}
\put(174,741){\scr$u$}
\put(198,771){\scr$u$}
\put(108,771){\scr$u$}
%\put( 60,768){\scr$-\!u\!-\!\lam$}\put(219,771){\scr$u$}
\put(14,764){\scr$K^{(1,2)}_+(u\!+\!\half\lam)$}
\put(230,764){\scr$K^{(1,2)}_-(u\!-\!\half\lam)$}
\end{picture}
\ee
which is $\T^{(2)}_0$. The another term with the
antisymmetric projector $Y^-$, which collapses
the matrix, is proportional to the identical matrix
\be
\setlength{\unitlength}{0.0125in}%
\begin{picture}(159,60)(56,375)
\put(105,375){\vector( 0, 1){ 60}}
\put(120,375){\vector( 0, 1){ 60}}
\put(195,375){\vector( 0, 1){ 60}}
\put(180,375){\vector( 0, 1){ 60}}
\multiput(138,405)(8.00000,0.00000){4}{\makebox(0.4444,0.6667){\tenrm .}}
\put( 36,396){}
\put(56,396){$f^1_0(u)$}
\end{picture}\ee
Therefore we have the equation (\ref{C1}). It can be seen that
the proof is correct also for the non-diagonal reflection $K$ matrices.
So it is  very clear that
the su($2$) fusion rule also works for the six-vertex or the eight-vertex
models with non-diagonal reflection matrices. These will be published
elsewhere.

\end{document}